\documentclass[traditabstract]{aa}
\usepackage{aalongtable}
\usepackage{natbib,twoopt}
 \bibpunct{(}{)}{;}{a}{}{,}    
\usepackage{graphics}
\usepackage{txfonts}
\usepackage{ulem}
\usepackage{times}
\usepackage{psfig}
\begin{document}
\title{New candidates for extremely metal-poor emission-line galaxies in the SDSS/BOSS DR10
}
\author{N. G. \ Guseva \inst{1,2}    
\and Y. I. \ Izotov \inst{1,2}
\and K. J. \ Fricke \inst{1,3}
\and C. \ Henkel \inst{1,4}
}
\offprints{N. G. Guseva, guseva@mao.kiev.ua}
\institute{          Max-Planck-Institut f\"ur Radioastronomie, Auf dem H\"ugel 
                     69, 53121 Bonn, Germany
\and
                     Main Astronomical Observatory,
                     Ukrainian National Academy of Sciences,
                     Zabolotnoho 27, Kyiv 03680,  Ukraine
\and 
                     Institut f\"ur Astrophysik, G\"ottingen Universit\"at, 
                     Friedrich-Hund-Platz 1, 37077 G\"ottingen, Germany 
\and
                     Astronomy Department, King Abdulaziz University, 
                     P.O.Box 80203, Jeddah 21589, Saudi Arabia
}
\date{Received \hskip 2cm; Accepted}

\abstract{
We present a spectroscopic study of eight extremely low-metallicity 
candidate emission-line galaxies
with oxygen abundances possibly below  
12 +log O/H = 7.35. These galaxies were
selected from Data Release 10 of the Sloan Digital Sky Survey/Baryon 
Oscillation Spectroscopic Survey (SDSS/BOSS DR10). 
We will call these extremely metal-deficient galaxies XMD galaxies.
The electron temperature-sensitive emission line
[O~{\sc iii}] $\lambda$4363 is detected in three galaxies
and marginally detected in two galaxies,
allowing  for abundance determination by a ``direct'' method. 
Because of large uncertainties in the [O {\sc iii}]$\lambda$4363\AA\ line
fluxes, we also calculated oxygen abundance in these galaxies together with
the remaining three galaxies using a strong-line semi-empirical method.
This method gives oxygen abundances higher than 7.35 for three galaxies with 
detected [O {\sc iii}]$\lambda$4363\AA\ line and lower than 7.35 for the 
remaining five objects of the sample.
The newly-discovered galaxies represent excellent targets for
follow-up spectroscopic observations with the largest telescopes to 
improve the oxygen abundance determination and to
increase the number of these very rare low-metallicity objects. 
The extreme location of the most massive and luminous XMD galaxies 
and XMD candidates in
the stellar mass-metallicity diagram implies that these galaxies may
be genuine young objects. 
 With stellar masses of up to $\sim$ 10$^7$ - 10$^8$$M_{\odot}$, 
the galaxies are not chemically enriched and
strongly deviate  to lower metallicity as compared to the relation 
obtained for a large sample of low-redshift, star-forming galaxies. 
}
\titlerunning{New XMD galaxies in SDSS/BOSS DR10}
\authorrunning{N. G. Guseva et al.}
\maketitle

\section {Introduction}

Searching for extremely metal-poor emission-line galaxies
is very important for several reasons. One of them is that these
galaxies are the most promising young galaxy candidates in the local Universe
\citep{G03,IT04b}.
In this paper we consider extremely metal-deficient (hereafter XMD) 
emission-line galaxies, searching for targets with oxygen abundances 
12 + logO/H $\leq$ 7.35.

The physical conditions in the interstellar medium (ISM) of XMDs 
provide insights into a variety of aspects of fundamental physics, cosmology,
and astronomy. 
 These galaxies have very low, but not pristine element abundances,
hinting at pre-enrichment by e.g.
Population III stars \citep{T05}. Large enough statistics is needed
to confirm this idea.
The galaxies with very low metallicity share many properties 
of Ly$\alpha$ emitting galaxies
and Ly-break galaxies \citep{P03,Nakajima2013,Shirazi2014,Shapley2014}. 
Thanks to the fact that XMDs can be excellent local proxies for their high-$z$ 
counterparts.
These chemically unevolved galaxies are also the best objects for the 
determination of the primordial $^4$He abundance $Y_p$ 
\citep[e.g. ][]{IT04a,ITS07,IT10,I2013,I2014b}. 
A more detailed review of their properties can be found in \citet{I12}.

Many efforts were undertaken recently to search for these very rare objects 
\citep[see e.g. ][]{P05,I06b,IT07,G07,IT09,Pustilnik2011,I12,B12}. 
More than ten galaxies with 12+log O/H $\leq$ 7.35 were found among the
galaxies from the Sloan Digital Sky Survey  
Data Release 7 (SDSS DR7) \citep{I12}.
 To further increase the number of these objects, we searched for new XMD
candidates among the objects from the extention of the SDSS survey, SDSS-III, 
namely among the galaxies that appeared in the SDSS/BOSS Data 
Release 10 (DR10).

%**********************************************

\setcounter{table}{0}

\begin{table*}
  \caption{General characteristics of newly identified XMD candidates \label{tab1}}
\begin{tabular}{lccccrccc} \hline \hline
Name&R.A.(J2000.0)$^{\rm a}$&Dec.(J2000.0)$^{\rm a}$&
redshift&$g$$^{\rm b}$&M$_g$$^{\rm c}$&M$_{\rm stars}$$^{\rm d}$&\multicolumn{2}{c}{12+logO/H$^{\rm e}$} \\
&&&&&&&T$_e$-method&strong-line method  \\  \hline
%&&&&Mpc&&& \\ \hline
%\multicolumn{7}{c}{a) 3.5m APO observations} \\ \hline
\object{J0100$-$0028} & 01:00:56.93 & $-$00:28:43.9 & 0.0192 & 20.85 & $-$14.43 & 6.62& ...  &7.10 \\ % ordered in R.A.
\object{J0122$+$0048} & 01:22:41.61 & $+$00:48:42.0 & 0.0574 & 21.62 & $-$16.16 & 6.45&  7.22 $\pm$ 0.15 &7.35 \\
\object{J0143$+$1958}$^{\rm f}$ & 01:43:15.15 & $+$19:58:06.1 & 0.0017 & 21.76 & $-$7.66& 4.28& ... &7.28 \\
\object{J0153$+$0104} & 01:53:11.96 & $+$01:04:40.1 & 0.0632 & 21.51 & $-$16.23 & 6.34&  7.34 $\pm$ 0.12 &7.60 \\
\object{J0222$-$0935} & 02:22:38.55 & $-$09:35:35.2 & 0.1148 & 21.61 & $-$17.30 & 8.04&  7.29 $\pm$ 0.32 &7.43 \\
\object{J0945$+$3835} & 09:45:19.55 & $+$38:35:52.9 & 0.0723 & 21.78 & $-$16.82 & 7.24&  7.33 $\pm$ 0.20 &7.48 \\
\object{J1036$+$2036} & 10:36:39.47 & $+$20:36:15.8 & 0.0549 & 21.59 & $-$15.61 & 6.82&  7.32 $\pm$ 0.34 &7.35 \\
\object{J1228$-$0125} & 12:28:45.54 & $-$01:25:26.9 & 0.0728 & 22.09 & $-$15.58 & 6.99& ...  & 7.33 \\ \hline %
\end{tabular}

$^{\rm a}$Equatorial coordinates. \\
$^{\rm b}$SDSS $g$ magnitude. \\
$^{\rm c}$Absolute $g$ magnitude. \\
$^{\rm d}$Log of stellar mass in unit of solar mass. \\
$^{\rm e}$This paper.\\ 
$^{\rm f}$J0143$+$1958 is part of galaxy \object{UGCA 20}.

  \end{table*}

%----------------------------------------------------------
\setcounter{figure}{0}

\begin{figure*}
\hbox{
\hspace{0.0cm}\psfig{figure=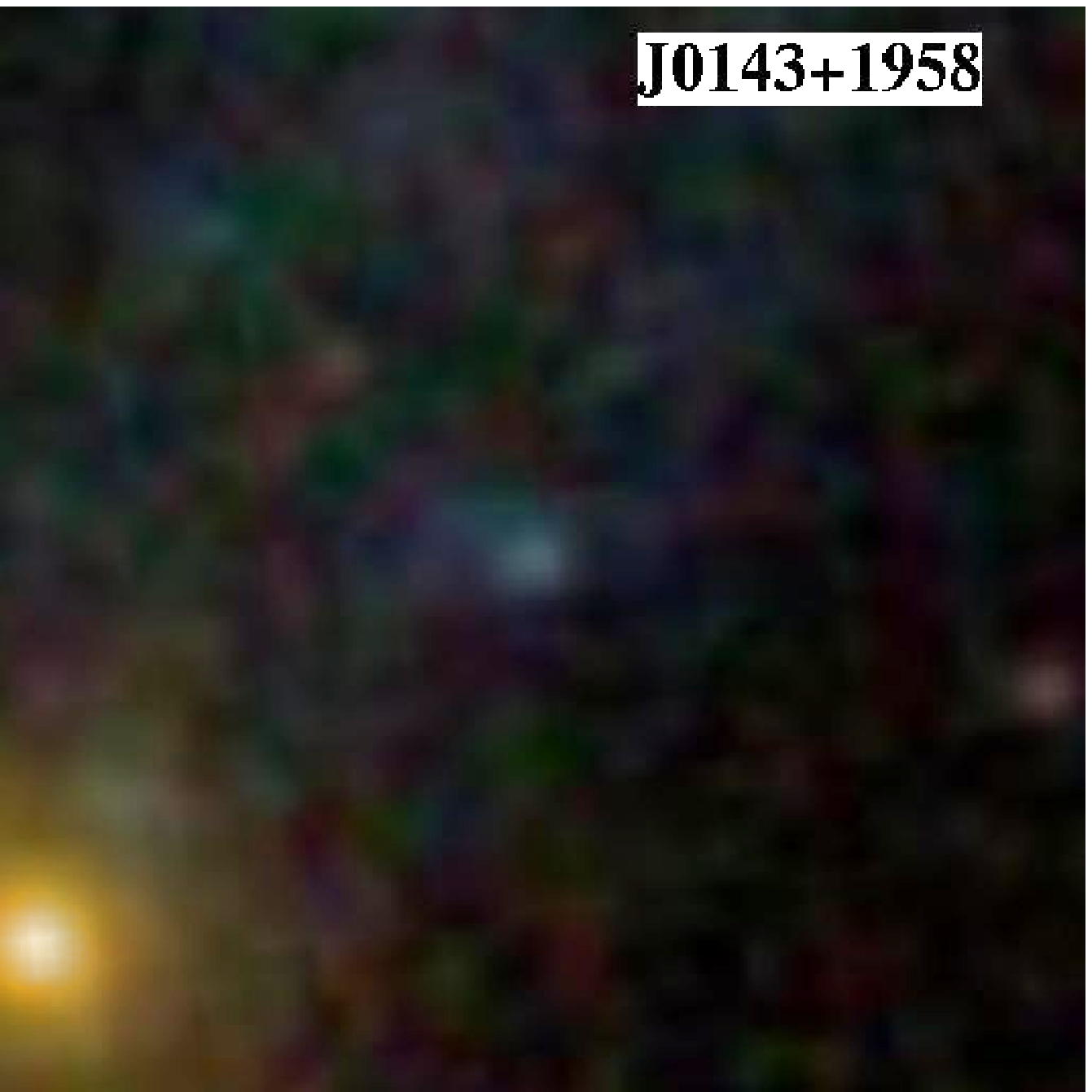,angle=0,width=4.5cm,clip=}
\hspace{0.0cm}\psfig{figure=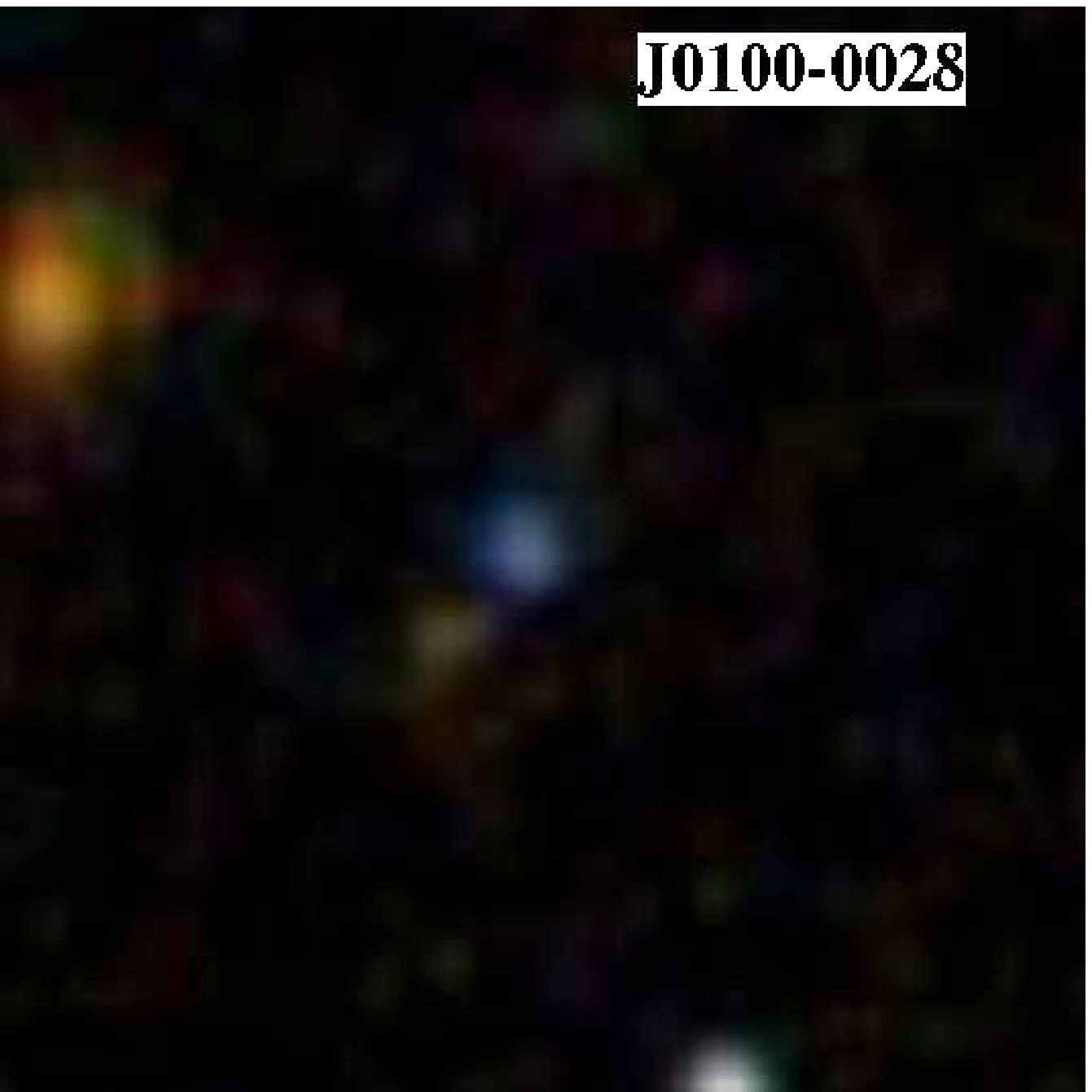,angle=0,width=4.5cm,clip=}
\hspace{0.0cm}\psfig{figure=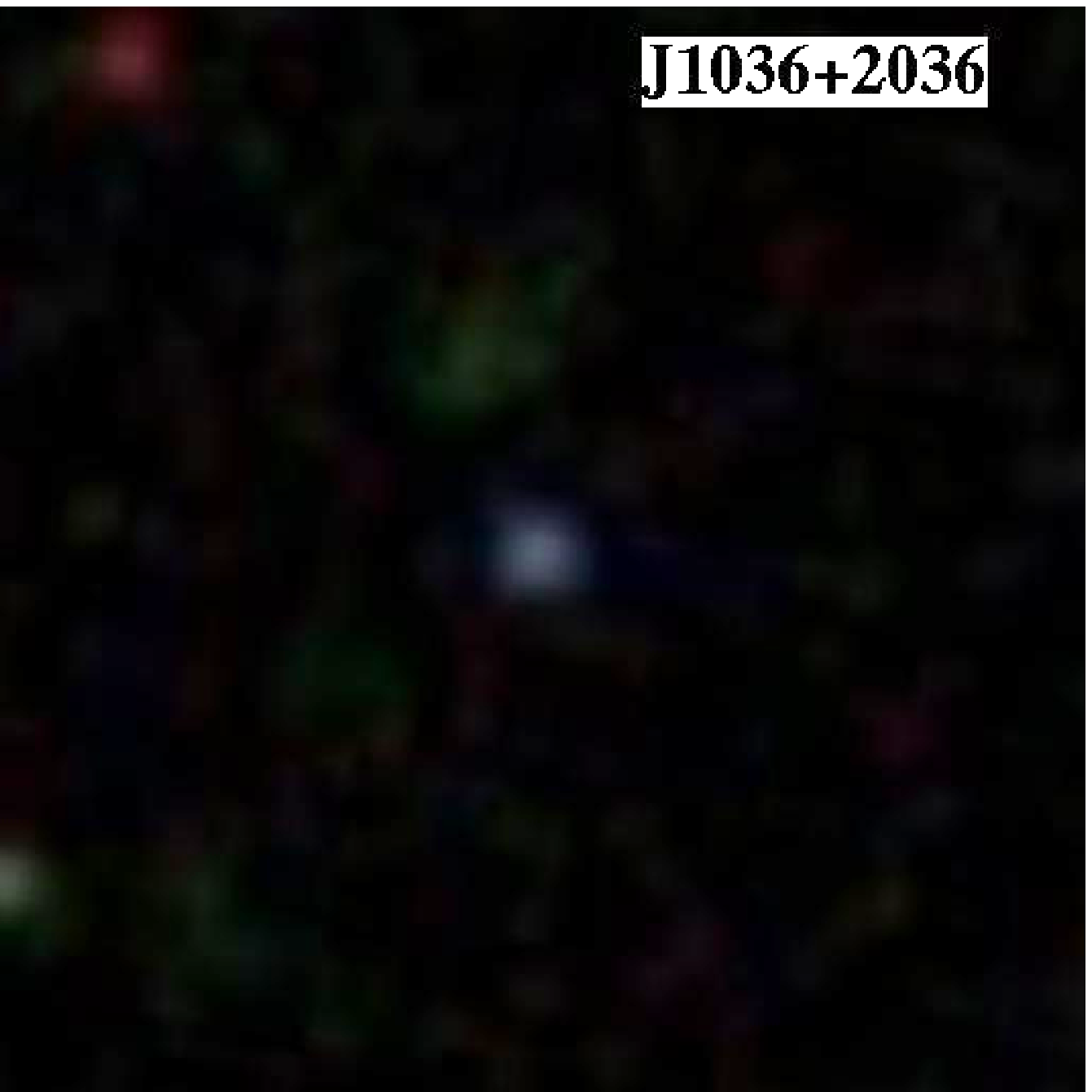,angle=0,width=4.5cm,clip=}
\hspace{0.0cm}\psfig{figure=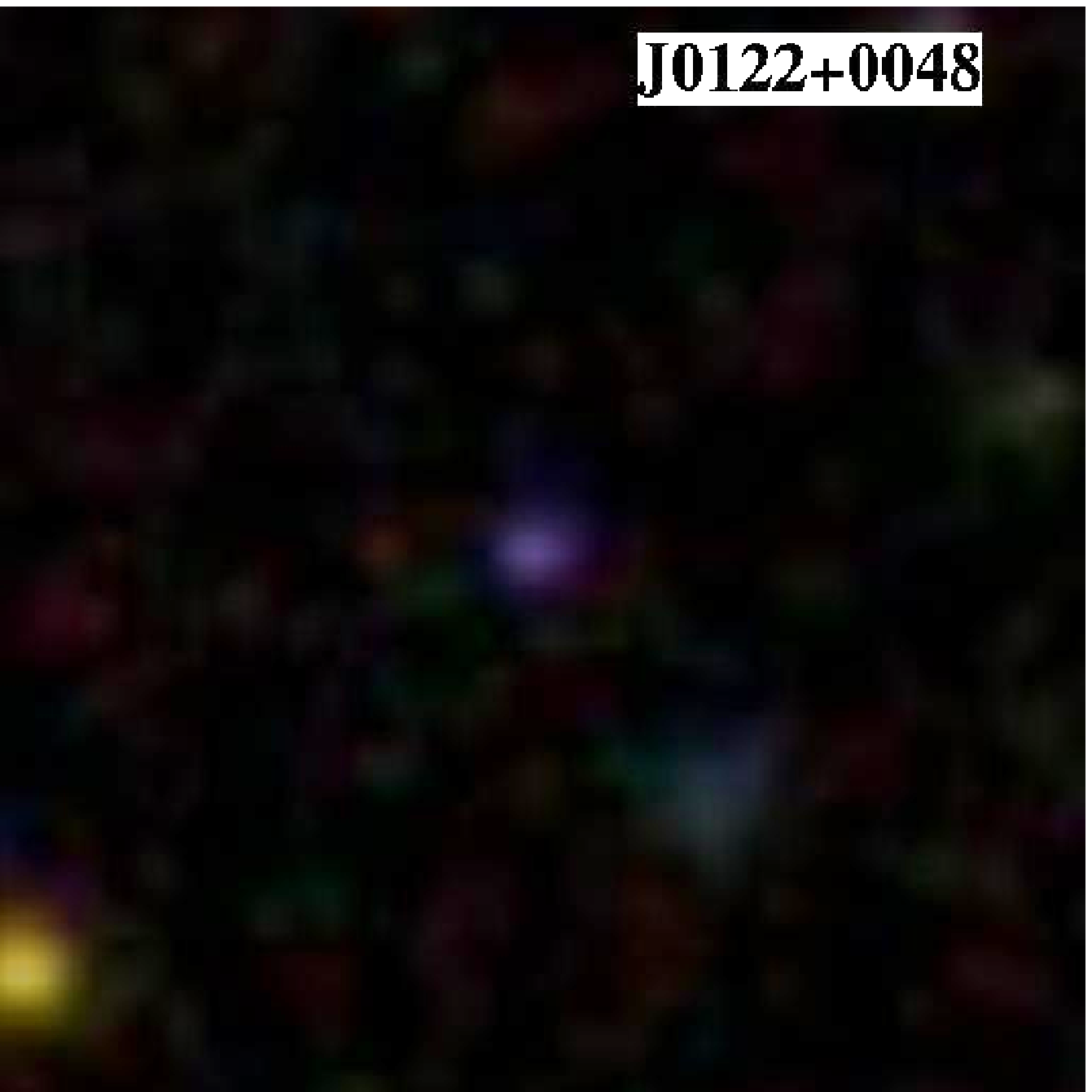,angle=0,width=4.5cm,clip=}
}
\vspace{0.1cm}
\hbox{
\hspace{0.0cm}\psfig{figure=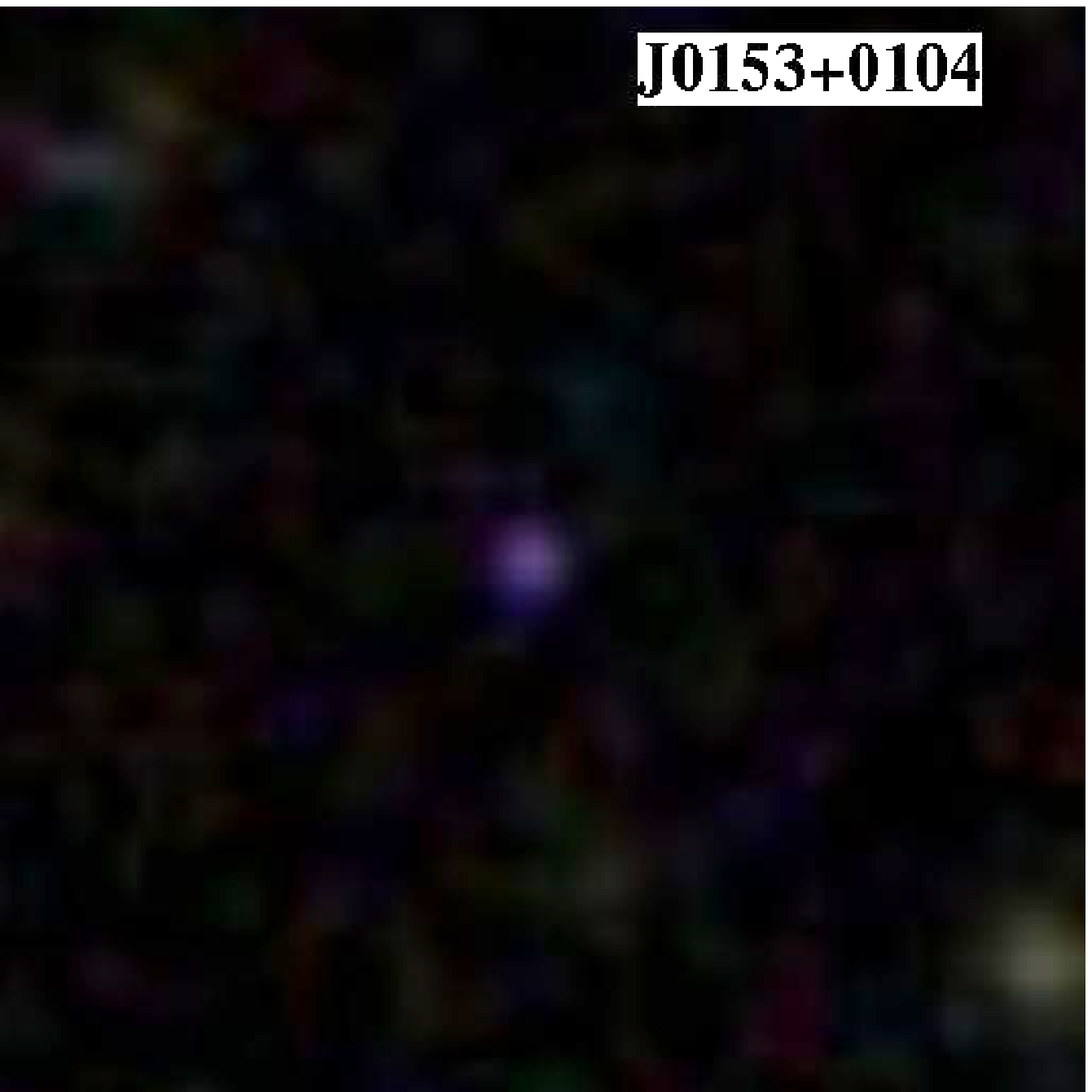,angle=0,width=4.5cm,clip=}
\hspace{0.0cm}\psfig{figure=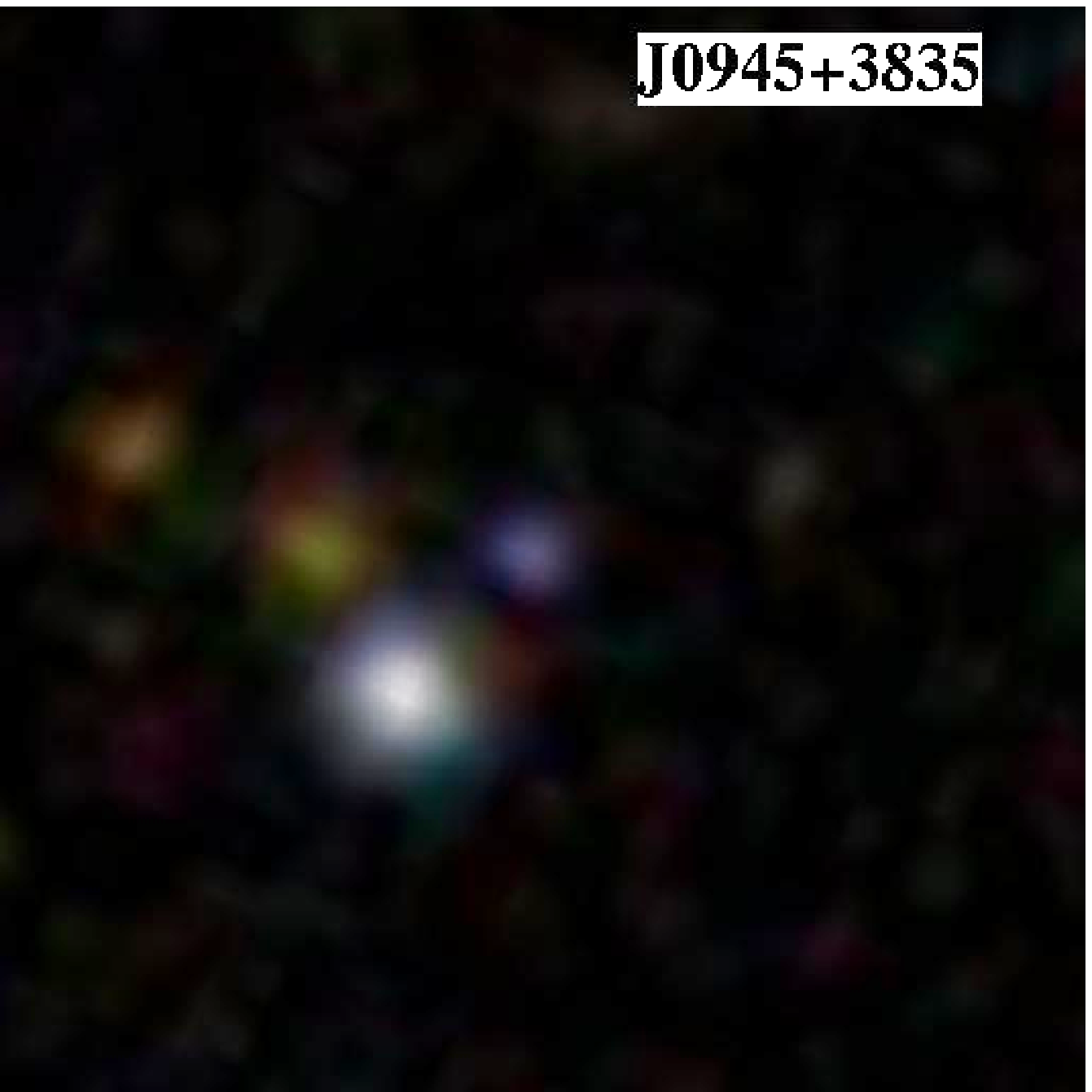,angle=0,width=4.5cm,clip=}
\hspace{0.0cm}\psfig{figure=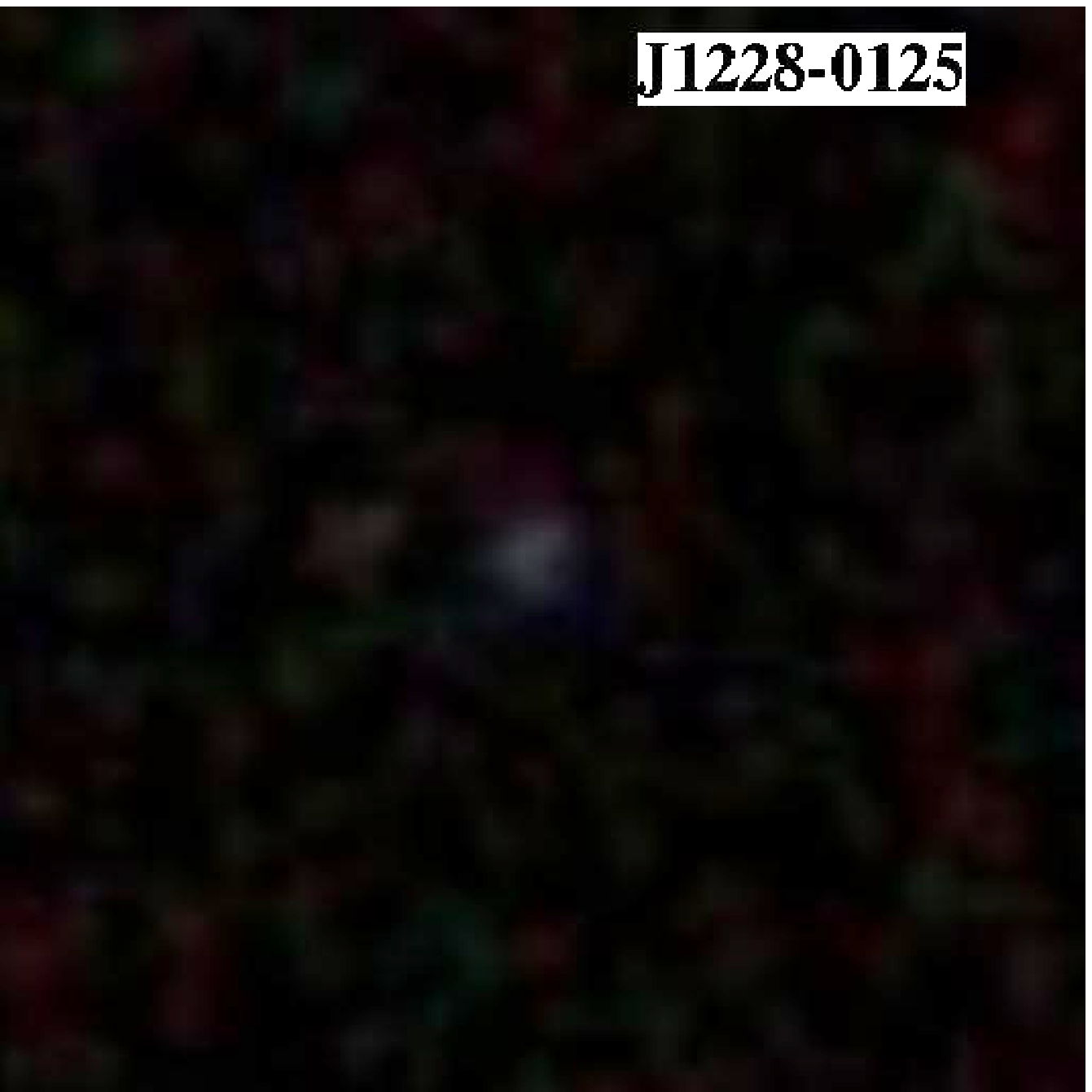,angle=0,width=4.5cm,clip=}
\hspace{0.0cm}\psfig{figure=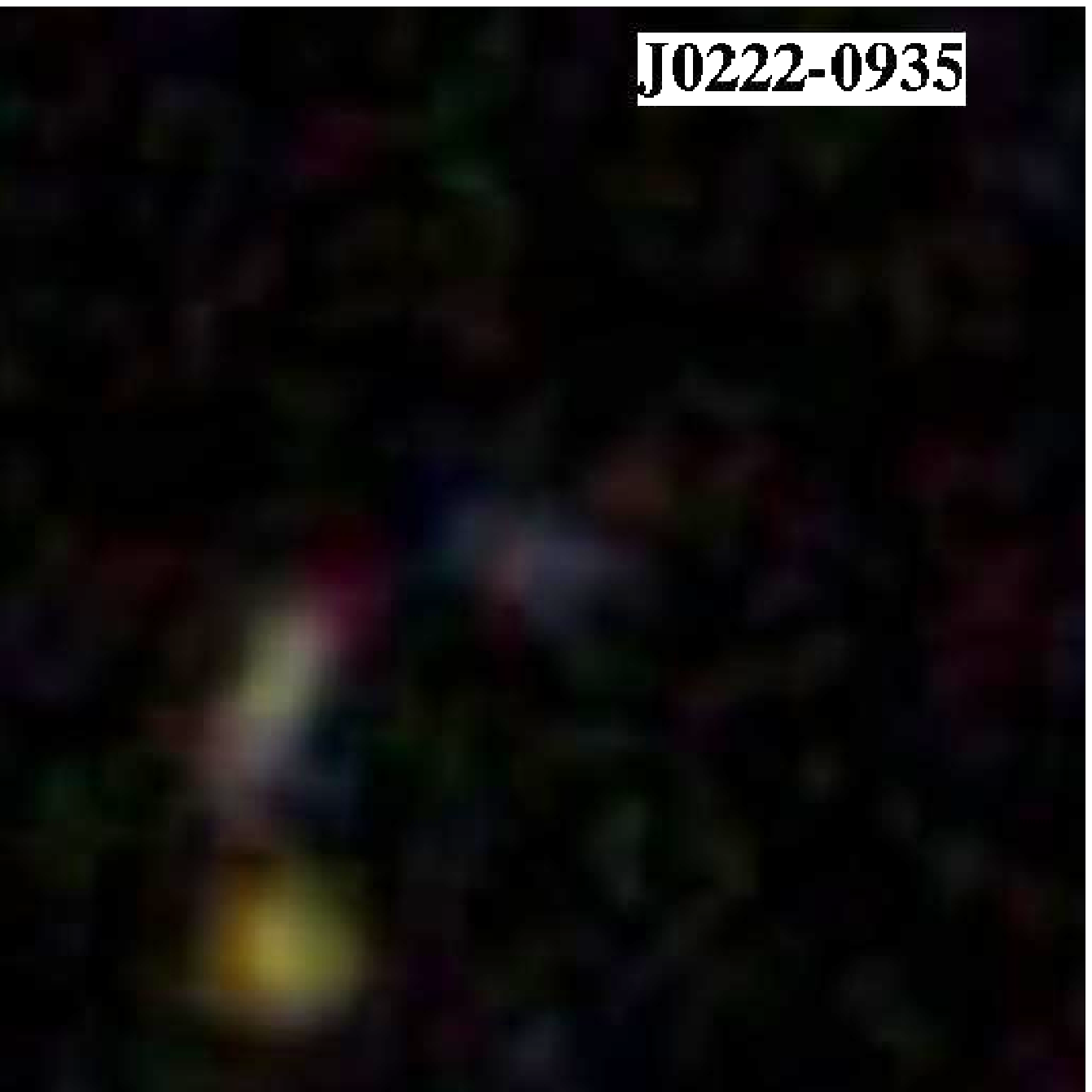,angle=0,width=4.5cm,clip=}
}
\caption{2\arcmin$\times$2\arcmin\ SDSS images of the galaxies ordered 
 according to increasing redshift.}
\label{fig1}
\end{figure*}

The data and the element abundance determination are described  in 
Section \ref{S2}.  
 The results are presented in Section \ref{S3}. More specifically,
we consider XMD emission-line galaxies  with respect to 
the luminosity metallicity relations
in Section \ref{S3s1}, their location in the emission-line diagnostic 
diagram in Section \ref{S3s2}, and in the mass-metallicity diagram in
Section \ref{S3s3}.  
Our main conclusions are summarised in Section \ref{S4}.

\section {The data \label{S2}}

 We use a sample of $\sim$9000 star-forming galaxies
by \citet{I2014a} selected from the spectroscopic database of the
SDSS/BOSS DR10 \citep{Ahn2014}. 
 The details of data selection can be found in \citet{I2014}. 
 Out of this large sample, we could only identify eight candidates 
that may represent extremely low-metallicity
 emission-line galaxies with 12 + logO/H $\leq$ 7.35. 
  The general characteristics of these galaxies, such as the name of galaxy, 
coordinates, redshift, apparent $g$ magnitude, absolute $g$ magnitude, 
and stellar mass are collected in Table \ref{tab1}. The last two columns
of the table show oxygen abundances we derived  
using two methods (see below).
 SDSS images of the galaxies with labelled names and ordered following 
increasing redshift are shown in Fig. \ref{fig1}.
 The rest-frame SDSS spectra of the galaxies arranged with increasing  
redshift are presented in Fig. \ref{fig2}.

 We measured the line fluxes and their errors using the 
IRAF\footnote{IRAF is the Image 
Reduction and Analysis Facility distributed by the National Optical Astronomy 
Observatory, which is operated by the Association of Universities for Research 
in Astronomy (AURA) under cooperative agreement with the National Science 
Foundation (NSF).} SPLOT routine.
  The line flux errors is later propagated into the calculation
of abundance errors.   
 We derived the internal extinction from the Balmer hydrogen emission lines 
after correction for the Milky Way extinction.
The line fluxes were corrected for both reddening \citep{W58}
and underlying hydrogen stellar absorption by the application of an 
iterative procedure \citep{ITL94}.
  The observed line intensities $F$($\lambda$)/$F$(H$\beta$),   
the extinction-corrected line intensities  $I$($\lambda$)/$I$(H$\beta$), 
multiplied by 100, 
and equivalent widths of emission lines are shown in Table \ref{tab3_1}
(available only in the online version of the paper) together with extinction 
coefficients  
$C$(H$\beta$), observed H$\beta$ fluxes, and equivalent widths $EW$(abs) of 
the hydrogen absorption stellar lines.

%----------------------------------------------------------
\setcounter{figure}{1}
\begin{figure*}
\hspace{0.0cm}\psfig{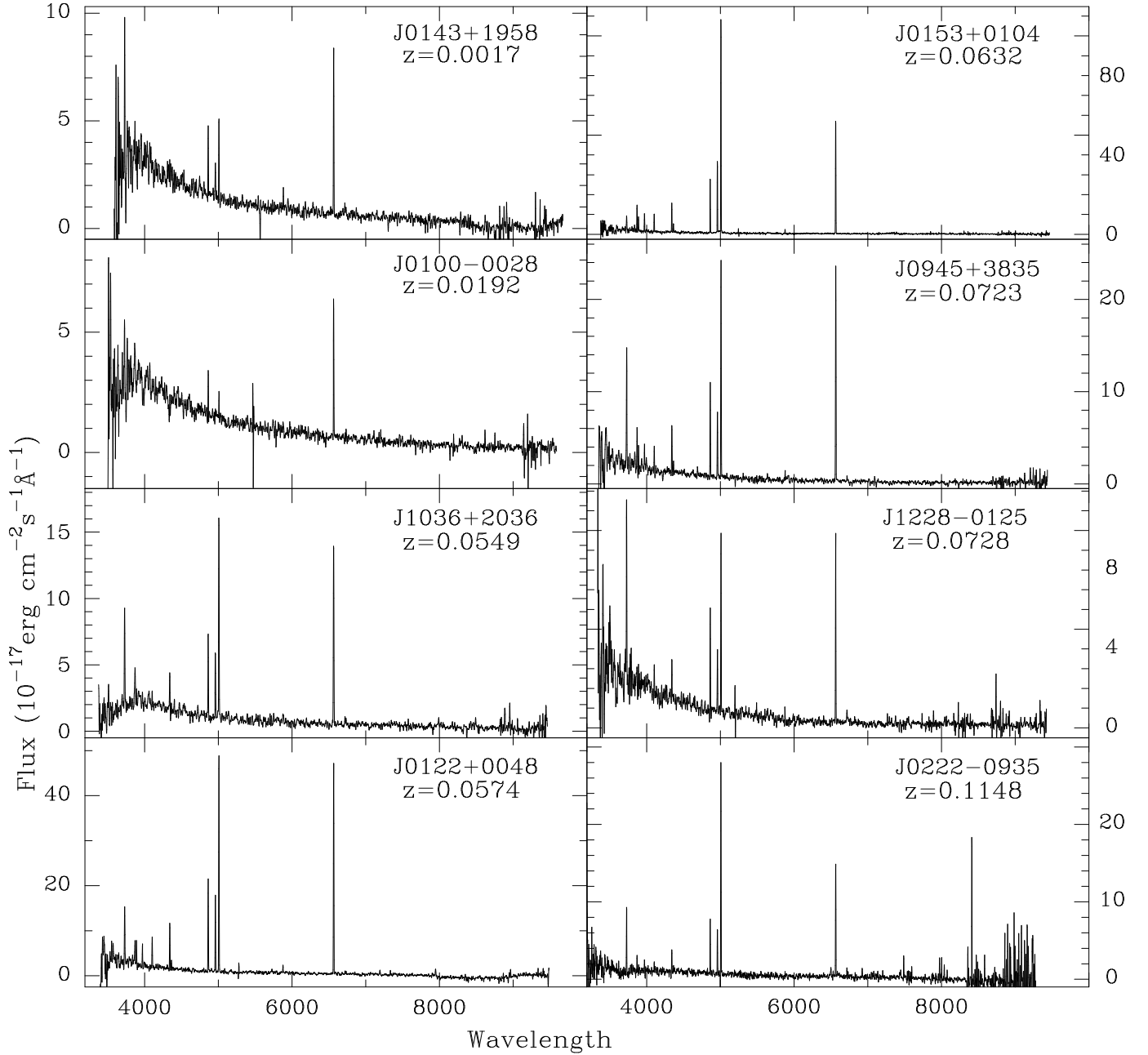}
\caption{SDSS spectra of the eight extremely metal-deficient galaxies.}
\label{fig2}
\end{figure*}
%----------------------------------------------------------

   Finally, we selected the five XMD candidates with the detected
[O {\sc iii}]$\lambda$4363\AA\ emission line.
This allows for a reliable oxygen abundance determination 
using the direct $T_{\rm e}$-method and following the prescriptions
of \citet{ITL94,ITL97} and \citet{TIL95}.
The temperature $T_{\rm e}$(O {\sc iii}) is calculated 
based on the [O {\sc iii}]$\lambda$4363/($\lambda$4959+$\lambda$5007) 
line ratio.
 For the remaining three XMD candidates the semi-empirical method by \citet{IT07} was 
applied to derive $T_{\rm e}$(O {\sc iii}). 
We adopt a two-zone photoionised H {\sc ii}
region model: a high-ionisation zone with temperature $T_{\rm e}$(O {\sc iii}), 
where [O {\sc iii}] and [Ne {\sc iii}] lines originate, and a 
low-ionisation zone with temperature $T_{\rm e}$(O {\sc ii}), where [O {\sc ii}]
lines originate.
   For $T_{\rm e}$(O {\sc ii}), we use
the relation between the electron temperatures $T_{\rm e}$(O {\sc iii}) and
$T_{\rm e}$(O {\sc ii}) obtained by \citet{I06} from
the H {\sc ii} photoionisation models of \citet{SI03}.

 We derived ionic and total oxygen and neon abundances  
using expressions for ionic abundances and the ionisation correction 
factor for neon 
by \citet{I06}. In Table \ref{tab3} (available only on line)
the electron temperatures $T_{\rm e}$(O {\sc ii}) and $T_{\rm e}$(O {\sc iii}), 
and oxygen and neon abundances are given 
for three galaxies with  
detected and for two galaxies with marginally detected [O {\sc iii}]4363 line. 
All five XMD candidates have very low oxygen
abundances 12 + log O/H $<$ 7.35 (Table \ref{tab3}). The oxygen abundance in 
the remaining three galaxies with undetected [O {\sc iii}]4363 lines is 
derived with the strong-line semi-empirical method (Table \ref{tab4}). 
These three
galaxies also have very low oxygen abundances 12 + log O/H $<$ 7.35.

Because of large uncertainties in the [O {\sc iii}]$\lambda$4363\AA\ line
fluxes, we also recalculated oxygen abundance for 
the five galaxies with detected and marginally detected 
[O {\sc iii}]$\lambda$4363\AA\ 
(Table \ref{tab3}) using a strong-line semi-empirical method. This 
method gives oxygen abundances higher than 7.35 for three galaxies with 
detected $\lambda$4363\AA\ line (see Table \ref{tab4}). 
The largest difference
is found for the galaxy \object{J0153+0104} with the best derived flux of the
[O {\sc iii}]$\lambda$4363\AA\ line and the best determination of the
oxygen abundance with the $T_{\rm e}$-method. This fact indicates that 
strong-line methods may not be
used for accurate abundance determination. Instead higher-quality observations
are needed to improve oxygen abundances in the considered galaxies.
We do not show the sulfur abundances
derived by using [S~{\sc iii}]$\lambda$9069 and $\lambda$9531 in 
Table \ref{tab3} and \ref{tab4}. These
lines are observed in noisy parts of the spectra with numerous residuals
of night sky lines, making sulfur abundances uncertain.

%---------------------------------------------------
\setcounter{figure}{2}

\begin{figure*}
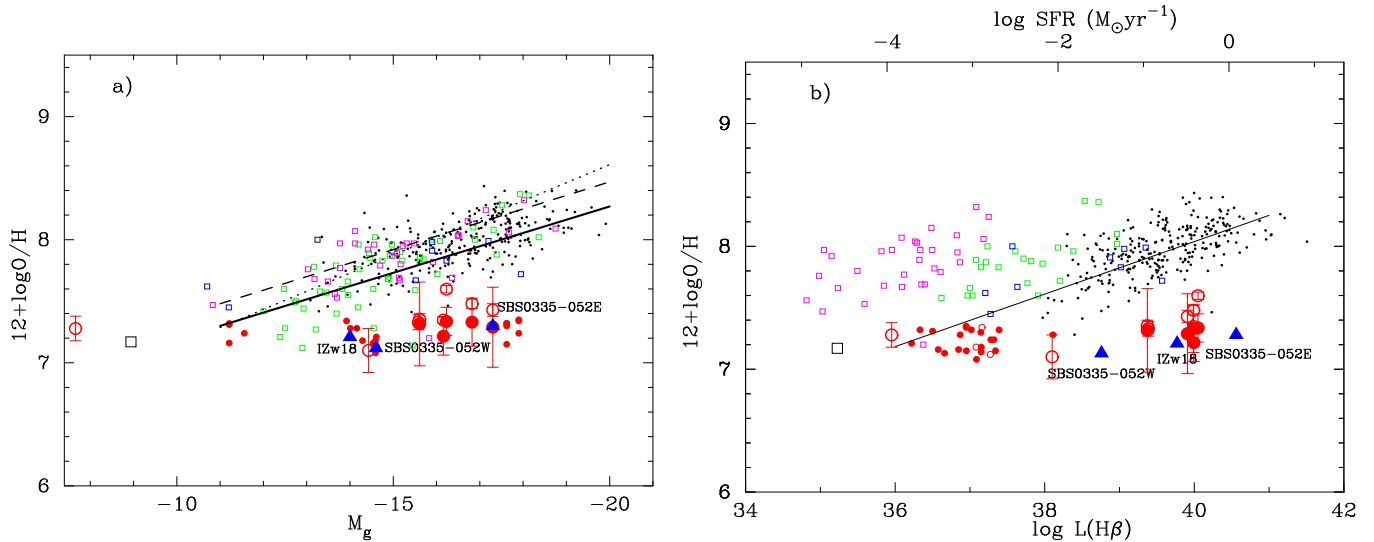

\hspace*{-0.1cm}\psfig{figure=fig_3a.ps,angle=-90,width=8.5cm,clip=}
\hspace*{0.25cm}\psfig{figure=fig_3b.ps,angle=-90,width=8.9cm,clip=}
\caption{(a) The absolute SDSS $g$ magnitude - oxygen abundance 
relation for the five galaxies with oxygen abundances derived using 
the ``direct''
method (large red filled circles) and for all eight galaxies with 
oxygen abundances derived with the semi-empirical method by \citet{IT07} 
(large red open circles).
 We also show the XMD emission-line galaxies with 12 + logO/H $\leq$ 7.35 
by \citet{IT07} and \citet{I12} 
in small red filled circles.
The small open squares in magenta, green, and blue are the data derived by
\citet{B12}, \citet{Pustilnik2011}, and \citet{James2014}, respectively 
 (see text).
  The most-metal deficient BCDs 
\object{I Zw 18}, \object{SBS 0335--052W}, and \object{SBS 0335--052E} 
are shown with filled blue triangles and are labelled. 
   One of the most metal-poor dwarf galaxies, \object{Leo P} from 
\citet{Skillman2013} is plotted with a large open black square.
  For comparison, the ``standard'' $L$ - $Z$ relation for emission-line 
galaxies \citep[SDSS and star-forming galaxies from ][]{I06,IT07}
are shown in black dots with the linear regression to these data
(solid line).
The relations derived by \citet{RM95} and by \citet{B12} are shown in a
dotted line and a dashed line, respectively.
(b) The extinction-corrected H$\beta$ luminosity - oxygen abundance relation. 
 The samples and symbols are the same as in (a).  The solid line represents 
the linear regression to SDSS galaxies \citep{I06} and to all galaxies 
from \citet{IT07}, excluding \object{I Zw 18}, \object{SBS 0335--052W}, and 
\object{SBS 0335--052E}, and to galaxies from \citet{I12}.
}
\label{fig3}
\end{figure*}

%**************************************************

\section{Results \label{S3}}

In this section, we discuss the oxygen abundances and the 
distributions of the extremely metal-deficient galaxies in the 
 broad-band luminosity-metallicity, H$\beta$ luminosity-metallicity,
diagnostic Baldwin-Phillips-Terlevich, and mass-metallicity diagrams.
 Throughout this Section we compare the XMD candidates with previous samples  
of similar emission-line galaxies, analysing all these data in a uniform 
way \citep{I06,IT07,I11,I12,I2014}. %method. %technics

It is seen from Table \ref{tab3} and \ref{tab4} that no 
galaxy with an oxygen abundance 12 + logO/H $<$ 7.1 was found.
 A similar conclusion was obtained in previous studies \citep[e.g. ][]{I12}.
This supports the idea
discussed by, e.g., 
\citet{T05} that the matter from which dwarf emission-line galaxies were
formed was pre-enriched to a level 12 + log O/H $\ga$ 6.9 \citep[or $\sim$
2\% of the abundance 12 + log O/H = 8.69 of the Sun, ][]{A09}.
They also showed that BCDs spanning a wide range in ionised gas metallicities 
all have  H {\sc i} envelopes with about the same 
neutral gas metallicity of $\sim$7.0. This is also the metallicity  
found in Ly$\alpha$ absorbers \citep{P03}. Taken together, the available data 
suggest that 
the primordial neutral gas may have been previously enriched to a common 
metallicity level 12 + log O/H $\sim$ 6.9 - 7.0, 
possibly by Population III stars.
 
 The number of XMDs with 
12 + logO/H $\leq$ 7.35 is yet very small. Only eight XMD candidates
out of $\sim$ 9000 emission-line
galaxies extracted from SDSS DR10, or $<$ 0.1\%, were 
found, indicating that
these objects are very scarce in the Local Universe.

\subsection{Luminosity-metallicity relations \label{S3s1}}

The existence of the relation between optical 
luminosities and metallicities of galaxies, the so called $L$ - $Z$ relation, 
was established 
by \citet{L79} and later confirmed by \citet{S89} and \citet{RM95}.
   The metallicity of a galaxy is a measure of the abundance of metals
relative to hydrogen in the interstellar medium and is most commonly
quoted in terms of the oxygen abundance 12 + log O/H. 
As a measure of the optical luminosity, we use the $M_g$ absolute magnitude 
obtained from the apparent SDSS $g$ magnitudes (Table \ref{tab1}), at variance
with early studies of the $L$ - $Z$ relation, in what the apparent $B$ magnitude
was commonly used. However, \citet{P08} have shown that SDSS $g$ magnitudes are
very similar to $B$ magnitudes of star-forming galaxies.
The distances to the galaxies are derived from their radial velocities, 
adopting a Hubble constant $H_0$ = 70 km s$^{-1}$ Mpc$^{-1}$.

  In Fig. \ref{fig3}a the absolute $g$ magnitude metallicity 
relation for SDSS galaxies by \citet{I06}, as a representative sample, 
is plotted in black dots. 
 We want to examine how much the XMDs 
differ from the star-forming galaxies in ``standard'' samples.
 The solid line represents the linear regression
to SDSS galaxies \citep{I06} and to all galaxies from \citet{IT07}. 
 The position of all XMD candidates 
from this paper and from \citet{IT07} and \citet{I12} 
are also plotted.

 \citet{B12} studied a sample of 42 low-luminosity galaxies aiming to
fill the relatively sparse low-luminosity end of the relation with
intrinsically faint galaxies.
 For comparison we show their data in small open squares in magenta. 
  Aiming to find the low-metallicity emission-line galaxies
in the voids \citet{Pustilnik2011} discovered 
some objects with low metallicity, which are plotted in Fig. \ref{fig3}
with green open squares.
 Blue diffuse dwarf (BDD) galaxies with star-formation activity from
\citet{James2014} are shown with open blue squares.
 The most-metal-deficient BCDs 
\object{I Zw 18}, \object{SBS 0335--052W}, and \object{SBS 0335--052E} are 
shown in filled blue triangles and are labelled. 
 One of the most metal-deficient star-forming nearby dwarf galaxies, Leo P
\citep{Skillman2013} is plotted with a large open black square.
  We also plotted the relation derived by \citet{RM95}
(dotted line) and the 
relation of \citet{B12} (dashed line),
which are not very different from the relation by \citet{IT07}.

  Most XMDs occupy the same region in the  $L - Z$ diagram as the well-known 
metal-deficient BCDs \object{I Zw 18} and \object{SBS 0335--052E}. All these 
galaxies are systematically shifted to lower 
metallicities relative to the “standard” line for
every bin of the $M_g$ range from $-$12 to $-$18 mag.
  Considering a higher oxygen abundance cutoff 12 + log(O/H) $\leq$ 7.65
\citet{K04a} and \citet{G09} found a large scatter of galaxies 
filling the gap between the solid line in Fig. \ref{fig3}a and the location
of XMDs with 12 + log(O/H) $<$ 7.35.

Several possible evolutionary scenarios were suggested for
these galaxies. 
 One of them implies that they are genuine chemically
unevolved (or young) galaxies with low metallicity.
At $M_g$ $\sim$ $-$17 they  are shifted down by about 0.5-0.6 dex 
relative to the “standard” $L-Z$ line.
  This implies that about 0.09\% of quite typical emission-line
galaxies, populating the Local Universe, show a sizable deficiency of
oxygen, amounting to not more than 2 - 4\% of the metallicity 
compared to the bulk of galaxies used for producing the 
$L-Z$ relation.

 A likely alternative interpretation of the observed shift
is an increase in luminosity at a given metallicity 
because of strong SF activity 
\citep[e.g. ][]{G09}.
  Most extreme XMD outliers are shifted from the ``standard'' $L-Z$ relation 
\citep[represented, for instance, with the relation of ][]{RM95} 
by more than 6-7 mags (Fig. \ref{fig3}a).
   Including most-metal deficient galaxies, especially those with strong and 
very strong ongoing star formation, leads to a shallower slope of 
the $L-M$ relation
\citep{G09}. In particular, XMDs with 12 + log(O/H) $<$ 7.35 and XMD 
candidates do not show a 
dependence on metallicity at all.

%---------------------------------------------
\setcounter{figure}{3}

\begin{figure}
\psfig{figure=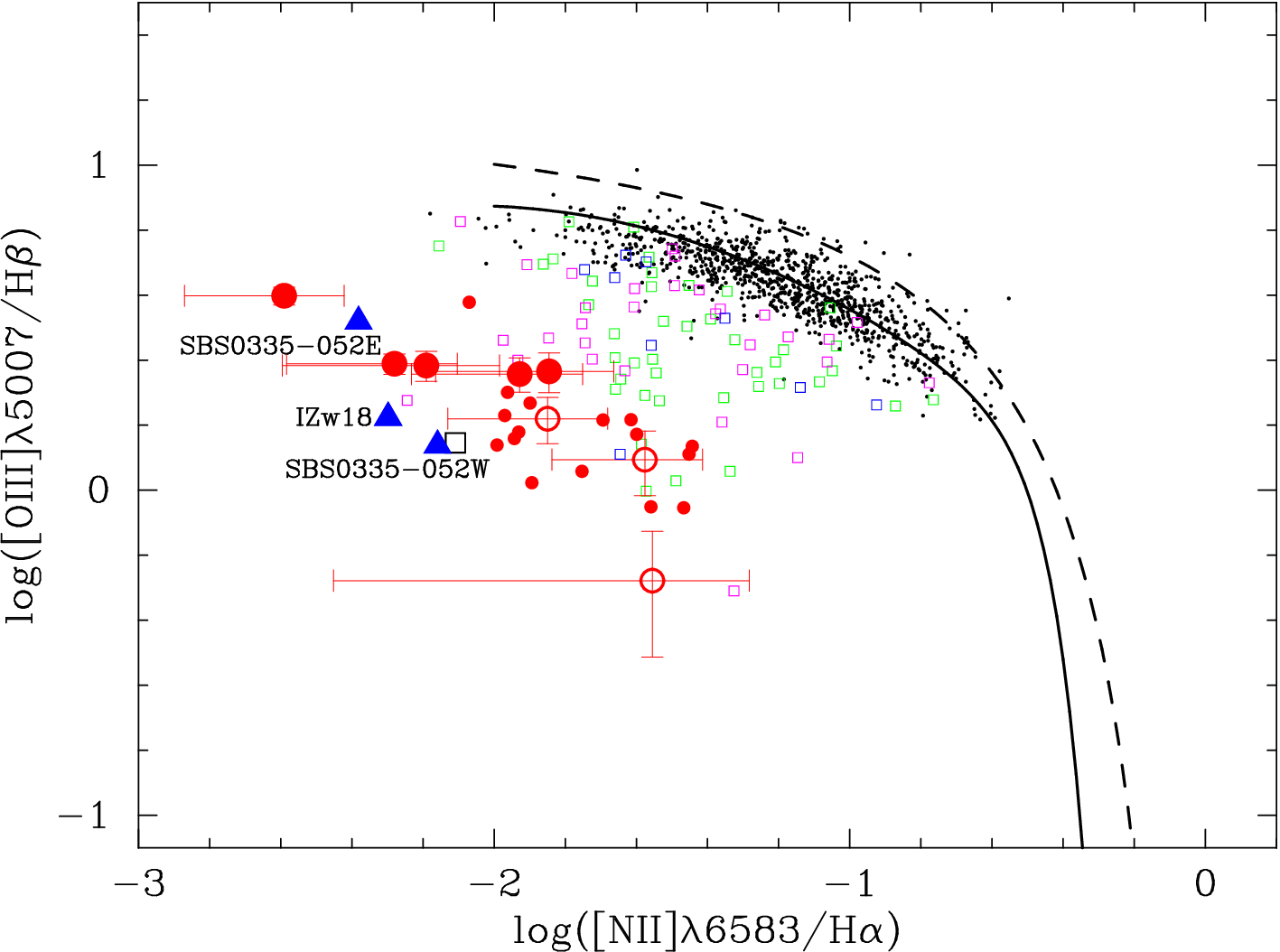,angle=-90,width=8.7cm,clip=}
\caption{The diagnostic diagram \citep[BPT, ][]{BPT81}
for the XMD emission-line galaxies with 12 + logO/H $\leq$ 7.35
and our XMD candidates.
Symbols are the same as in Fig. \ref{fig3}.
 We also plot a sample of 803 luminous compact galaxies 
\citep[LCGs, black dots, ][]{I11}.
 The dashed line from \citet{K03} and the solid
line from \citet{S06} separate star-forming galaxies (left side) from active 
galactic nuclei (right side). 
}
\label{fig4}
\end{figure}

%***************************************************

   In Fig. \ref{fig3}b, the diagram H$\beta$ luminosity 
$L$(H$\beta$) metallicity is plotted for the same galaxies as in 
Fig. \ref{fig3}a. The upper axis 
is a scale for the star formation
rate (SFR), derived from the H$\alpha$ luminosity as defined by
\citet{K98}.
  As seen in Fig. \ref{fig3}b, XMDs have a large spread
of H$\beta$ luminosity from 10$^{36}$ to 10$^{40}$ erg s$^{-1}$. 
  Six out of eight galaxies from our new XMD candidate sample 
are characterised by strong SF activity, which results in the
presence of high-excitation H {\sc ii} regions with strong emission lines.
  Their SFRs are in the range 
$\sim$0.01 - 0.5 $M_\odot$ yr$^{-1}$ and are close to the SFRs of 
\object{I Zw 18} and the \object{SBS 0335--052} system. 
  Only one object from our newly discovered XMDs, \object{J0143+1958}, has an
H$\beta$ luminosity corresponding to one late O or early B star 
assuming that a single O5V star produces the
H$\beta$ luminosity of 
4.8$\times$10$^{36}$ erg s$^{-1}$ \citep[leftmost outlying open circles in Figs. \ref{fig3}a and \ref{fig3}b, ][]{L90}.
%(leftmost outlying open circles in Figs. \ref{fig3}a and \ref{fig3}b).
This is in fact not a galaxy, but an H~{\sc ii} region in the nearby
low-surface-brightness dwarf galaxy \object{UGCA 20}, while our other XMDs are 
more distant compact
galaxies with spectra, which are representative for the entire galaxy.

 The galaxies by \citet{B12} denoted by small open magenta squares
in Fig. \ref{fig3}a have log $L$(H$\beta$) in
the range of 35 -- 37 and belong to galaxies with low-excitation 
H {\sc ii} regions
and higher metallicity as compared to our sample of XMD galaxies and XMD 
candidates. 

  Thus, intense starbursts with high SFRs, similarly to the 
well-known BCDs \object{I Zw 18} and\object{SBS 0335--052E} \citep{SS70,I90},
were discovered in six out of eight XMD candidates.

\subsection{The emission-line diagnostic diagram\label{S3s2}}

 In Fig. \ref{fig4} we plot the location of XMDs 
in the Baldwin-Phillips-Terlevich (BPT) diagnostic 
diagram \citep{BPT81}. 
 As seen in the figure, the newly
discovered most metal-poor galaxies are located at the same place as the XMDs
with 12 + logO/H $\leq$ 7.35 collected by \citet{I12}. 
All these galaxies are 
far below the region, where the
luminous compact galaxies (LCGs) by \citet{I11} are located (black dots). 
Solid and dashed lines are the lines by \citet{S06} and 
\citet{K03}, respectively, separating star-forming galaxies and AGNs.
Similar to Fig. \ref{fig3} the location of most newly discovered 
XMD candidates 
in Fig. \ref{fig4} is close to that of the most metal-deficient BCDs
\object{I Zw 18} and \object{SBS 0335--052E+W}.
The XMDs presently known populate a region that is so far almost 
empty. Thus, our discovery of these very scarce objects in the 
Local Universe appreciably increases
the statistics needed to understand the nature of these galaxies.

\setcounter{figure}{4}

\begin{figure}
\psfig{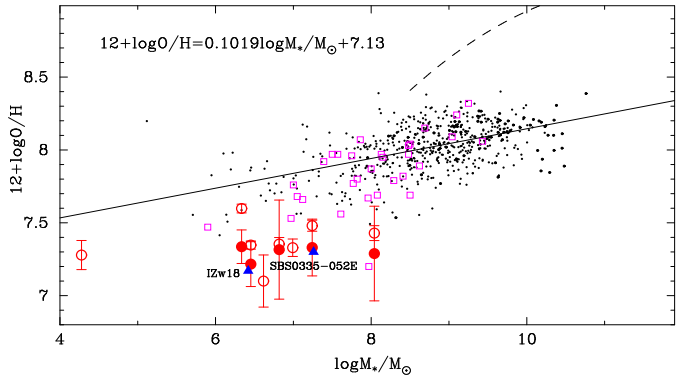}
\caption{The stellar mass oxygen abundance relation. Emission-line XMD
galaxies with 12 + log O/H $<$ 7.35 and our XMD candidates  
are shown in red and blue.
 Symbols are the same as in Fig. \ref{fig3}.
 For comparison, the SDSS DR7 emission-line galaxy sample by \citet{I2014}, 
where the errors in [O~{\sc iii}]$\lambda$4363 emission-line flux do not 
exceed 25\%, are shown in black dots. The linear likelihood regression 
for this SDSS DR7 sample is plotted with a solid line. 
The small open squares in magenta are the data derived by \citet{B12}.
 The mass-metallicity relation obtained from 53000 SDSS star-forming galaxies 
by \citet{Tremonti2004} is plotted with a dashed line.
 }
\label{fig5}
\end{figure}

\subsection{Mass-metallicity relation \label{S3s3}}

The relation between the stellar mass and its metallicity is a
fundamental relation linking these two global galaxy properties.
  Mass is
more fundamentally related to metallicity than luminosity.
The luminosity of a galaxy is enhanced (active phase) or not enhanced (passive
phase) by ongoing star formation, while stellar mass and metallicity are
both determined  by a galaxy's evolution in the past.
  We show our XMD candidates on the stellar mass-metallicity 
($M_*$-$Z$) diagram
(Fig. \ref{fig5}). The solid line is the $M_*$-$Z$
relation obtained by \citet{I2014} for SDSS DR7 emission-line galaxies  
with 
robust metallicity determinations (black dots). 
For these galaxies, the errors of the [O~{\sc iii}] 4363 
emission-line fluxes do not exceed 25\%.

To derive stellar masses we modelled spectral energy distributions (SEDs) 
in the entire spectral range of $\lambda$$\lambda$3600-10300\AA, 
and subtracted gaseous line and continuum emission. 
 The method is based on fitting a series of model SEDs to
the observed SED and finding the best fit \citep{G06,G07,I11}.
 Star formation history was approximated 
by a recent short burst in ages $<$ 10 Myr, which accounts for
the young stellar population, and a prior continuous star formation
responsible for older stars of ages, that vary between 
10 Myr and 15 Gyr.

 Our eight XMD candidates are shown in large red circles. 
All of them, except the
low-mass H {\sc ii} region in \object{UGCA 20}, are located in the same area, as
\object{I Zw 18} and \object{SBS 0335--052E}. They are shifted by several 
orders 
of magnitude to high stellar masses as compared to stellar masses
defined by the $M_*$-$Z$ relation for SDSS DR7 emission-line galaxies. 
For the range of stellar masses of these galaxies of 
$\sim$2$\times$10$^6$ - 10$^8$$M_{\odot}$, 
their oxygen abundances are 3-4 times lower than predicted
from  the $M_*$-$Z$ relation for the 
SDSS DR7 emission-line galaxy sample.
These differences may likely be due to the relatively young ages 
of XMDs, which had not enough time to evolve to higher metallicities, 
and support the idea that these %galaxies are
least chemically evolved galaxies are genuine young galaxies.
  Many XMD galaxies are interacting systems 
\citep{Ekta2008}. \citet{Ekta2010} proposed an alternative hypothesis that XMD
galaxies deviate from the $L$-$Z$ and $M_*$-$Z$ relations because of a 
combination of having low effective chemical yields and higher gas fractions
due to tidal interactions.

Presently, XMDs are characterised by very high specific star formation
rates (sSFR) of 10 -- 100 Gyr$^{-1}$. These high sSFRs contradict 
the conclusion by
\citet{Shi2014} that the star formation efficiences in dwarf low-metallicity
galaxies are less than a tenth of the one found in normal metal-rich galaxies
and that therefore star formation may have been inefficient in the early
Universe.

\section{Conclusions \label{S4}}

We present a spectroscopic study of eight candidates for extremely 
metal-deficient (XMD) emission-line galaxies %newly 
selected  
from the Sloan Digital Sky Survey/Baryon Oscillation Spectroscopic Survey 
Data Release 10 (SDSS/BOSS DR10), a part of the SDSS-III survey. %database 
 We merged this sample with analogous XMD galaxies from the SDSS DR7
\citep{I12} and from previous investigations. Our main results are as 
follows.
 
     We find that oxygen abundances in all eight galaxies are
very low and classify them as XMD candidates. 
The oxygen abundances in five XMD candidates with detected or marginally 
detected
[O {\sc iii}]$\lambda$4363 emission line were derived by the ``direct'' method.
Note, however, the large uncertainties in the weak  
[O~{\sc iii}]$\lambda$4363 line in some galaxies.
Six out of eight newly discovered XMD candidates are characterised 
by strong SF activity and high SFRs, similar to those in the well-known 
BCDs \object{I Zw 18} and \object{SBS 0335--052E} \citep{SS70,I90}. 

  No emission-line galaxies with 12 + logO/H $\sim$ 7.0 were found
in the entire combined SDSS DR7 and DR10 surveys and among other galaxies 
collected from the literature \citep[this paper and ][]{I12}.
This finding supports conclusions  
by \citet{T05} and \citet{I12} that the matter, from which XMDs were formed,
was probably pre-enriched to this level prior to their formation. 

  The XMDs with strong SF activity are the most prominent outliers
in the luminosity-metallicity ($L-Z$) and 
H$\beta$ luminosity-metallicity ($L$(H$\beta$) - $Z$) relations,
increasing the dispersion in the low-metallicity end of the relations.

   Extreme positions of the most massive and luminous XMDs in
the mass-metallicity diagrams imply that the galaxies could be
genuinely young. These galaxies with stellar masses  in the range of
$\sim$ 10$^7$ - 10$^8$ $M_{\odot}$ are not yet chemically enriched.

\acknowledgements

 Y.I.I. and N.G.G. thank the hospitality of the Max-Planck 
Institute for Radioastronomy, Bonn, Germany.    
    Funding for the Sloan Digital Sky Survey (SDSS-III) has been 
provided by the Alfred P. Sloan Foundation, the Participating Institutions, 
the National Science Foundation, the U.S. Department of Energy, the National 
Aeronautics and Space Administration, the Japanese Monbukagakusho, and the 
Max Planck Society, and the Higher Education Funding Council for England. 
 
%------------------------------------------------------------

\Online

\setcounter{table}{1}

\begin{table*}
%\tiny
%\small
\caption{Emission line intensities and       equivalent widths}
\label{tab3_1}
\begin{tabular}{lrrrcrrr} \hline
&\multicolumn{7}{c}{\sc Galaxy} \\  \cline{2-8}     
&
 \multicolumn{3}{c}{          \object{J0143+1958}}&&
 \multicolumn{3}{c}{          \object{J0100--0028}} \\ 
  \cline{2-4} \cline{6-8} 
  {Ion}
  &{$F$($\lambda$)/$F$(H$\beta$)}
  &{$I$($\lambda$)/$I$(H$\beta$)}
  &{EW$^{a}$}&
  &{$F$($\lambda$)/$F$(H$\beta$)}
  &{$I$($\lambda$)/$I$(H$\beta$)}
  &{EW$^{a}$} \\ \hline

3727 [O {\sc ii}]                 & 159.1 $\pm$  36.7 & 141.8 $\pm$  39.0 &  15.7 & &  94.2 $\pm$  31.2 & 106.6 $\pm$  37.7 &   8.0 \\
4340 H$\gamma$                    &  37.0 $\pm$  16.6 &  46.9 $\pm$  25.0 &   6.1 & &  42.7 $\pm$  19.5 &  46.7 $\pm$  24.0 &   6.6 \\
4861 H$\beta$                     & 100.0 $\pm$  25.1 & 100.0 $\pm$  28.7 &  20.7 & & 100.0 $\pm$  30.0 & 100.0 $\pm$  32.2 &  14.9 \\
4959 [O {\sc iii}]                &  48.6 $\pm$  17.4 &  43.1 $\pm$  17.4 &   8.9 & &  19.0 $\pm$  18.9 &  18.4 $\pm$  18.8 &   1.8 \\
5007 [O {\sc iii}]                & 139.9 $\pm$  31.1 & 124.2 $\pm$  31.1 &  23.3 & &  54.7 $\pm$  22.5 &  52.7 $\pm$  22.2 &   6.6 \\
6563 H$\alpha$                    & 318.1 $\pm$  62.1 & 287.9 $\pm$  68.8 & 113.8 & & 343.1 $\pm$  79.8 & 288.3 $\pm$  74.8 &  69.7 \\
6716 [S {\sc ii}]                 &  19.1 $\pm$  11.2 &  16.9 $\pm$  11.3 &   5.1 & &  17.8 $\pm$  13.4 &  14.7 $\pm$  11.5 &   4.0 \\
6731 [S {\sc ii}]                 &   9.9 $\pm$  10.0 &   8.8 $\pm$  10.0 &   2.7 & &  21.9 $\pm$  14.7 &  18.1 $\pm$  12.6 &   4.3 \\
9531 [S {\sc iii}]                &   \multicolumn{1}{c}{...} &   \multicolumn{1}{c}{...} &   ... & &  47.8 $\pm$  16.5 &  34.1 $\pm$  14.0 &  29.1 \\
 $C$(H$\beta$) & \multicolumn{3}{c}{ 0.005 }& & \multicolumn{3}{c}{ 0.205 } \\
 $F$(H$\beta$)$^{b}$ & \multicolumn{3}{c}{  1.40 }& & \multicolumn{3}{c}{  0.97 } \\
 EW(abs)$^{a}$ \AA & \multicolumn{3}{c}{ 2.61 }& & \multicolumn{3}{c}{ 0.33 } \\ \hline

\end{tabular}
\end{table*}

\setcounter{table}{1}

\begin{table*}
%\tiny
%\small
\caption{\it Continued}
%\label{tab3_2}
\begin{tabular}{lrrrcrrr} \hline
&\multicolumn{7}{c}{\sc Galaxy} \\  \cline{2-8}
&
 \multicolumn{3}{c}{          \object{J1036+2036}}&&
 \multicolumn{3}{c}{          \object{J0122+0048}} \\
  \cline{2-4} \cline{6-8} 
{Ion}
  &{$F$($\lambda$)/$F$(H$\beta$)}
  &{$I$($\lambda$)/$I$(H$\beta$)}
  &{EW$^{a}$}&
  &{$F$($\lambda$)/$F$(H$\beta$)}
  &{$I$($\lambda$)/$I$(H$\beta$)}
  &{EW$^{a}$} \\ \hline

3727 [O {\sc ii}]                 &  88.2 $\pm$  13.5 &  91.5 $\pm$  14.7 &  31.3 & &  48.7 $\pm$   5.7 &  57.0 $\pm$   6.9 &  29.0 \\
3869 [Ne {\sc iii}]               &  19.5 $\pm$   7.6 &  20.1 $\pm$   7.9 &   5.2 & &  21.5 $\pm$   3.8 &  24.6 $\pm$   4.4 &  18.3 \\
3889 He {\sc i} + H8              &   \multicolumn{1}{c}{...} &   \multicolumn{1}{c}{...} &   ... & &  23.7 $\pm$   3.9 &  27.5 $\pm$   5.9 &  24.2 \\
3968 [Ne {\sc iii}] + H7          &   \multicolumn{1}{c}{...} &   \multicolumn{1}{c}{...} &   ... & &  25.4 $\pm$   3.9 &  29.0 $\pm$   5.4 &  31.9 \\
4101 H$\delta$                    &  26.8 $\pm$   6.8 &  28.0 $\pm$   8.2 &  20.2 & &  27.6 $\pm$   4.1 &  30.9 $\pm$   5.5 &  33.0 \\
4340 H$\gamma$                    &  46.8 $\pm$   8.9 &  48.1 $\pm$  10.5 &  27.6 & &  47.7 $\pm$   5.3 &  51.2 $\pm$   6.2 &  78.5 \\
4363 [O {\sc iii}]                &   7.3 $\pm$   5.6 &   7.4 $\pm$   5.7 &   3.2 & &   9.2 $\pm$   2.8 &   9.8 $\pm$   3.0 &  10.3 \\
4861 H$\beta$                     & 100.0 $\pm$  14.1 & 100.0 $\pm$  14.7 &  78.9 & & 100.0 $\pm$   8.8 & 100.0 $\pm$   9.1 & 147.5 \\
4959 [O {\sc iii}]                &  72.5 $\pm$  11.6 &  71.9 $\pm$  11.6 &  32.5 & &  86.7 $\pm$   7.9 &  85.5 $\pm$   7.8 & 179.6 \\
5007 [O {\sc iii}]                & 229.9 $\pm$  27.5 & 227.6 $\pm$  27.4 &  91.3 & & 249.8 $\pm$  18.5 & 244.8 $\pm$  18.3 & 468.7 \\
5876 He {\sc i}                   &   \multicolumn{1}{c}{...} &   \multicolumn{1}{c}{...} &   ... & &  11.2 $\pm$   2.5 &  10.0 $\pm$   2.3 &  22.7 \\
6300 [O {\sc i}]                  &   \multicolumn{1}{c}{...} &   \multicolumn{1}{c}{...} &   ... & &   1.4 $\pm$   1.6 &   1.2 $\pm$   1.3 &   3.2 \\
6312 [S {\sc iii}]                &   \multicolumn{1}{c}{...} &   \multicolumn{1}{c}{...} &   ... & &   0.6 $\pm$   1.5 &   0.5 $\pm$   1.3 &   1.5 \\
6563 H$\alpha$                    & 287.8 $\pm$  33.2 & 273.9 $\pm$  34.6 & 238.2 & & 324.5 $\pm$  23.4 & 272.6 $\pm$  21.5 &1015.0 \\
6716 [S {\sc ii}]                 &  13.4 $\pm$   4.9 &  12.7 $\pm$   4.7 &  11.3 & &   4.0 $\pm$   1.8 &   3.3 $\pm$   1.5 &   9.4 \\
6731 [S {\sc ii}]                 &  10.0 $\pm$   4.5 &   9.5 $\pm$   4.3 &   8.3 & &   3.8 $\pm$   1.7 &   3.1 $\pm$   1.4 &   9.0 \\
9069 [S {\sc iii}]                &   \multicolumn{1}{c}{...} &   \multicolumn{1}{c}{...} &   ... & &   1.8 $\pm$   1.1 &   1.3 $\pm$   0.8 &  17.9 \\
9531 [S {\sc iii}]                &   \multicolumn{1}{c}{...} &   \multicolumn{1}{c}{...} &   ... & &  20.7 $\pm$   3.1 &  14.6 $\pm$   2.4 &  56.6 \\
 $C$(H$\beta$) & \multicolumn{3}{c}{ 0.060 }& & \multicolumn{3}{c}{ 0.225 } \\
 $F$(H$\beta$)$^{b}$ & \multicolumn{3}{c}{  3.10 }& & \multicolumn{3}{c}{  8.10 } \\
 EW(abs)$^{a}$ \AA & \multicolumn{3}{c}{ 0.44 }& & \multicolumn{3}{c}{ 0.41 } \\ \hline

\end{tabular}
\end{table*}

\setcounter{table}{1}

\begin{table*}
%\tiny
%\small
\caption{\it Continued}
%\label{tab3_2}
\begin{tabular}{lrrrcrrr} \hline
&\multicolumn{7}{c}{\sc Galaxy} \\   \cline{2-8}
&
 \multicolumn{3}{c}{          \object{J0153+0104}}&&
 \multicolumn{3}{c}{          \object{J0945+3835}} \\
  \cline{2-4} \cline{6-8}
  {Ion}
  &{$F$($\lambda$)/$F$(H$\beta$)}
  &{$I$($\lambda$)/$I$(H$\beta$)}
  &{EW$^{a}$}&
  &{$F$($\lambda$)/$F$(H$\beta$)}
  &{$I$($\lambda$)/$I$(H$\beta$)}
  &{EW$^{a}$} \\ \hline

3727 [O {\sc ii}]                 &  23.3 $\pm$   3.5 &  25.6 $\pm$   4.0 &  34.1 & & 103.1 $\pm$  12.9 & 129.9 $\pm$  17.2 &  41.2 \\
3869 [Ne {\sc iii}]               &  38.0 $\pm$   4.5 &  41.1 $\pm$   5.0 &  47.0 & &  35.0 $\pm$   6.6 &  42.7 $\pm$   8.3 &  19.7 \\
3889 He {\sc i} + H8              &  20.1 $\pm$   3.4 &  22.4 $\pm$   5.5  &  24.6 & &  32.6 $\pm$   5.8 &  39.7 $\pm$   7.6 &  57.2 \\
3968 [Ne {\sc iii}] + H7         &  29.6 $\pm$   3.9 &  32.2 $\pm$   5.2 &  49.8 & &  31.2 $\pm$   6.1 &  37.5 $\pm$   9.3 &  22.9 \\
4101 H$\delta$                    &  26.7 $\pm$   3.6 &  28.7 $\pm$   4.8 &  48.3 & &  26.9 $\pm$   5.5 &  31.4 $\pm$   7.8 &  25.0 \\
4340 H$\gamma$                    &  46.5 $\pm$   4.9 &  48.6 $\pm$   5.6 &  98.2 & &  50.0 $\pm$   7.7 &  55.4 $\pm$   9.7 &  42.8 \\
4363 [O {\sc iii}]                &  14.5 $\pm$   2.8 &  15.0 $\pm$   2.9 &  31.2 & &  10.1 $\pm$   3.9 &  11.1 $\pm$   4.3 &   9.4 \\
4686 He {\sc ii}                  &   \multicolumn{1}{c}{...} &   \multicolumn{1}{c}{...} &   ... & &   6.3 $\pm$   3.7 &   6.5 $\pm$   3.8 &   4.6 \\
4861 H$\beta$                     & 100.0 $\pm$   8.2 & 100.0 $\pm$   8.5 & 215.5 & & 100.0 $\pm$  12.2 & 100.0 $\pm$  12.9 &  87.6 \\
4959 [O {\sc iii}]                & 132.5 $\pm$  10.2 & 131.1 $\pm$  10.2 & 180.7 & &  86.4 $\pm$  10.5 &  84.7 $\pm$  10.4 &  79.4 \\
5007 [O {\sc iii}]                & 401.9 $\pm$  26.7 & 396.3 $\pm$  26.5 & 582.4 & & 248.8 $\pm$  25.5 & 241.9 $\pm$  25.0 & 246.3 \\
5876 He {\sc i}                   &   9.9 $\pm$   2.1 &   9.0 $\pm$   2.0 &  41.6 & &  14.5 $\pm$   3.8 &  12.3 $\pm$   3.3 &  27.2 \\
6563 H$\alpha$                    & 303.1 $\pm$  20.6 & 272.5 $\pm$  20.3 &1432.0 & & 350.1 $\pm$  34.5 & 271.5 $\pm$  29.3 & 440.3 \\
6716 [S {\sc ii}]                 &   1.8 $\pm$   1.2 &   1.6 $\pm$   1.1 &  10.0 & &   \multicolumn{1}{c}{...} &   \multicolumn{1}{c}{...} &   ... \\
6731 [S {\sc ii}]                 &   1.8 $\pm$   1.2 &   1.6 $\pm$   1.1 &   9.2 & &   \multicolumn{1}{c}{...} &   \multicolumn{1}{c}{...} &   ... \\
9069 [S {\sc iii}]                &   4.1 $\pm$   1.3 &   3.4 $\pm$   1.2 &  38.2 & &   \multicolumn{1}{c}{...} &   \multicolumn{1}{c}{...} &   ... \\
 $C$(H$\beta$) & \multicolumn{3}{c}{ 0.135 }& & \multicolumn{3}{c}{ 0.330 } \\
 $F$(H$\beta$)$^{b}$ & \multicolumn{3}{c}{  9.35 }& & \multicolumn{3}{c}{  3.98 } \\
 EW(abs)$^{a}$ \AA & \multicolumn{3}{c}{ 0.75 }& & \multicolumn{3}{c}{ 0.22 } \\ \hline

\end{tabular}
\end{table*}

\setcounter{table}{1}

\begin{table*}
%\tiny
%\small
\caption{\it Continued}
%\label{tab3_2}
\begin{tabular}{lrrrcrrr} \hline
&\multicolumn{7}{c}{\sc Galaxy} \\  \cline{2-8}
 &
 \multicolumn{3}{c}{          \object{J1228--0125}}&&
 \multicolumn{3}{c}{          \object{J0222--0935}} \\
  \cline{2-4} \cline{6-8}
  {Ion}
  &{$F$($\lambda$)/$F$(H$\beta$)}
  &{$I$($\lambda$)/$I$(H$\beta$)}
  &{EW$^{a}$}&
  &{$F$($\lambda$)/$F$(H$\beta$)}
  &{$I$($\lambda$)/$I$(H$\beta$)}
  &{EW$^{a}$} \\ \hline

3727 [O {\sc ii}]                 & 130.1 $\pm$  22.7 & 127.5 $\pm$  23.9 &  23.3 & & 109.9 $\pm$  17.4 & 108.1 $\pm$  19.3 &  83.5 \\
3869 [Ne {\sc iii}]               &  12.6 $\pm$   9.9 &  12.3 $\pm$   9.9 &   2.8 & &  24.3 $\pm$   7.7 &  23.8 $\pm$   8.1 &  15.7 \\
3889 He {\sc i} + H8              &   \multicolumn{1}{c}{...} &   \multicolumn{1}{c}{...} &   ... & &  15.1 $\pm$   6.1 &  17.8 $\pm$   7.9 &  16.7 \\
3968 [Ne {\sc iii}] + H7          &   \multicolumn{1}{c}{...} &   \multicolumn{1}{c}{...} &   ... & &  16.4 $\pm$   7.2 &  23.3 $\pm$  11.7 &   7.5 \\
4101 H$\delta$                    &  30.0 $\pm$  10.5 &  35.5 $\pm$  13.9 &   9.6 & &  16.0 $\pm$   6.6 &  20.4 $\pm$   9.4 &  10.8 \\
4340 H$\gamma$                    &  39.9 $\pm$  11.3 &  44.7 $\pm$  13.8 &  14.1 & &  39.9 $\pm$   9.3 &  42.9 $\pm$  10.9 &  28.0 \\
4363 [O {\sc iii}]                &   \multicolumn{1}{c}{...} &   \multicolumn{1}{c}{...} &   ... & &  10.0 $\pm$   6.4 &   9.6 $\pm$   6.5 &   4.8 \\
4861 H$\beta$                     & 100.0 $\pm$  17.5 & 100.0 $\pm$  17.9 &  99.8 & & 100.0 $\pm$  16.3 & 100.0 $\pm$  17.6 &  53.4 \\
4959 [O {\sc iii}]                &  56.8 $\pm$  12.9 &  55.6 $\pm$  12.9 &  22.1 & &  86.0 $\pm$  14.6 &  80.5 $\pm$  14.6 &  50.6 \\
5007 [O {\sc iii}]                & 169.3 $\pm$  26.4 & 166.0 $\pm$  26.4 &  65.1 & & 248.3 $\pm$  33.8 & 232.1 $\pm$  33.7 &  73.7 \\
6563 H$\alpha$                    & 264.7 $\pm$  37.8 & 261.0 $\pm$  41.3 & 340.8 & & 299.7 $\pm$  39.4 & 271.1 $\pm$  41.3 & 284.3 \\
6716 [S {\sc ii}]                 &   \multicolumn{1}{c}{...} &   \multicolumn{1}{c}{...} &   ... & &  23.6 $\pm$   6.7 &  21.0 $\pm$   6.5 &  22.6 \\
6731 [S {\sc ii}]                 &   \multicolumn{1}{c}{...} &   \multicolumn{1}{c}{...} &   ... & &  12.5 $\pm$   5.4 &  11.1 $\pm$   5.2 &  10.2 \\
 $C$(H$\beta$) & \multicolumn{3}{c}{ 0.000 }& & \multicolumn{3}{c}{ 0.065 } \\
 $F$(H$\beta$)$^{b}$ & \multicolumn{3}{c}{  2.08 }& & \multicolumn{3}{c}{  2.41 } \\
 EW(abs)$^{a}$ \AA & \multicolumn{3}{c}{ 2.00 }& & \multicolumn{3}{c}{ 3.45 } \\ \hline
\end{tabular}

$^{a}$In angstroms. \\
$^{b}$In units of 10$^{-16}$ ergs s$^{-1}$ cm$^{-2}$.
\end{table*}

\setcounter{table}{2}

\begin{table*}
%\tiny
%\small
\caption{Ionic and total heavy element abundances (T$_e$-method)}
\label{tab3}
\begin{tabular}{lrrrrr} \hline
  &\multicolumn{5}{c}{\sc Galaxy} \\  \hline       
 {\sc Property}&
\multicolumn{1}{c}{        \object{J1036+2036}} & \multicolumn{1}{c}{        \object{J0122+0048}} &
\multicolumn{1}{c}{        \object{J0153+0104}} & \multicolumn{1}{c}{        \object{J0945+3835}} &
\multicolumn{1}{c}{        \object{J0222-0935}} \\ \hline

  $T_{\rm e}$(O {\sc iii}) (K) &
$19694 \pm 9122 $ & $22129 \pm 4638 $ &
$21544 \pm 2824 $ & $24350 \pm 7071 $ & $22604 \pm 10766 $\\

  $T_{\rm e}$(O {\sc ii}) (K) &
$15583 \pm 6684 $ & $15540 \pm 2928 $ &
$15604 \pm 1859 $ & $14985 \pm 3595 $ & $15462 \pm 6547 $ \\
  \\
  \\
  O$^+$/H$^+$ ($\times$10$^4$) &
$ 0.074\pm 0.081$ & $ 0.049\pm 0.024$ &
$ 0.022\pm 0.007$ & $ 0.118\pm 0.076$ & $ 0.089\pm 0.097$ \\
  O$^{++}$/H$^+$ ($\times$10$^4$) &
$ 0.132\pm 0.140$ & $ 0.116\pm 0.053$ &
$ 0.195\pm 0.057$ & $ 0.096\pm 0.059$ & $ 0.105\pm 0.109$ \\
  O/H ($\times$10$^4$) &
$ 0.207\pm 0.162$ & $ 0.165\pm 0.058$ &
$ 0.217\pm 0.057$ & $ 0.214\pm 0.096$ & $ 0.195\pm 0.146$ \\
  12 + log(O/H)  &
$ 7.316\pm 0.340$ & $ 7.216\pm 0.154$ &
$ 7.336\pm 0.115$ & $ 7.330\pm 0.195$ & $ 7.289\pm 0.325$ \\
  \\
  Ne$^{++}$/H$^+$ ($\times$10$^5$) &
$ 0.264\pm 0.279$ & $ 0.249\pm 0.111$ &
$ 0.440\pm 0.126$ & $ 0.354\pm 0.198$ & $ 0.230\pm 0.223$ \\
  ICF &
  1.153 &  1.125 &
  1.043 &  1.243 &  1.198 \\
  log(Ne/O) &
$-0.833\pm 0.709$ & $-0.769\pm 0.288$ &
$-0.674\pm 0.176$ & $-0.686\pm 0.479$ & $-0.849\pm 0.727$ \\
  \\ \hline

\end{tabular}
\end{table*}

\setcounter{table}{3}

\begin{table*}
%\tiny
%\small
\caption{Ionic and total heavy element abundances (strong-line method)}
\label{tab4}
\begin{tabular}{lrrrr} \hline
  &\multicolumn{4}{c}{\sc Galaxy} \\  \hline       
 {\sc Property}&
\multicolumn{1}{c}{        \object{J0143+1958}} & \multicolumn{1}{c}{        \object{J0100--0028}} &
\multicolumn{1}{c}{        \object{J1036+2036}} & \multicolumn{1}{c}{        \object{J0122+0048}} \\ \hline

  $T_{\rm e}$(O {\sc iii}) (K) &
$19549 \pm 1016 $ & $22841 \pm 1592 $ &
$18307 \pm 516 $ & $18214 \pm 323 $  \\

  $T_{\rm e}$(O {\sc ii}) (K) &
$15567 \pm 1278 $ & $15416 \pm 2119 $ &
$15344 \pm 635 $ & $15321 \pm 397 $ \\
  \\
  \\
  O$^+$/H$^+$ ($\times$10$^4$) &
$ 0.115\pm 0.039$ & $ 0.102\pm 0.051$ &
$ 0.073\pm 0.014$ & $ 0.049\pm 0.007$ \\
  O$^{++}$/H$^+$ ($\times$10$^4$) &
$ 0.075\pm 0.018$ & $ 0.023\pm 0.010$ &
$ 0.152\pm 0.018$ & $ 0.174\pm 0.013$ \\
  O/H ($\times$10$^4$) &
$ 0.190\pm 0.044$ & $ 0.126\pm 0.052$ &
$ 0.226\pm 0.023$ & $ 0.222\pm 0.014$ \\
  12 + log(O/H)  &
$ 7.279\pm 0.100$ & $ 7.100\pm 0.179$ &
$ 7.354\pm 0.045$ & $ 7.347\pm 0.028$ \\
  \\
  Ne$^{++}$/H$^+$ ($\times$10$^5$) &
  \multicolumn{1}{c}{...}& \multicolumn{1}{c}{...} &
$ 0.297\pm 0.120$ & $ 0.374\pm 0.069$ \\
  ICF &
  \multicolumn{1}{c}{...} &  \multicolumn{1}{c}{...} &
  1.138 &  1.092 \\
  log(Ne/O) &
 \multicolumn{1}{c}{...}&  \multicolumn{1}{c}{...} &
$-0.825\pm 0.233$ & $-0.736\pm 0.098$ \\
  \\ \hline
  &\multicolumn{4}{c}{\sc Galaxy} \\  \hline       
  {\sc Property}&
\multicolumn{1}{c}{        \object{J0153+0104}} & \multicolumn{1}{c}{        \object{J0945+3835}} &
\multicolumn{1}{c}{        \object{J1228--0125}} & \multicolumn{1}{c}{        \object{J0222--0935}} \\ \hline
  $T_{\rm e}$(O {\sc iii}) (K) &
$16067 \pm 308 $ & $17307 \pm 417 $ &
$18827 \pm  645 $ & $17690 \pm 579 $  \\
  $T_{\rm e}$(O {\sc ii}) (K) &
$14553 \pm 366 $ & $15052 \pm 506 $ &
$15456 \pm  802 $ & $15176 \pm 707 $  \\
  \\
  \\
  O$^+$/H$^+$ ($\times$10$^4$) &
$ 0.026\pm 0.004$ & $ 0.110\pm 0.017$ &
$ 0.106\pm 0.024$ & $ 0.091\pm 0.019$ \\
  O$^{++}$/H$^+$ ($\times$10$^4$) &
$ 0.370\pm 0.027$ & $ 0.192\pm 0.019$ &
$ 0.108\pm 0.017$ & $ 0.177\pm 0.025$ \\
  O/H ($\times$10$^4$) &
$ 0.396\pm 0.027$ & $ 0.302\pm 0.026$ &
$ 0.213\pm 0.029$ & $ 0.269\pm 0.032$ \\
  12 + log(O/H)  &
$ 7.598\pm 0.030$ & $ 7.480\pm 0.038$ &
$ 7.329\pm 0.060$ & $ 7.429\pm 0.051$ \\
  \\
  Ne$^{++}$/H$^+$ ($\times$10$^5$) &
$ 0.868\pm 0.112$ & $ 0.732\pm 0.148$ &
$ 0.180\pm 0.145$ & $ 0.395\pm 0.135$ \\
  ICF &
  1.026 &  1.156 &
  1.215 &  1.145 \\
  log(Ne/O) &
$-0.648\pm 0.066$ & $-0.552\pm 0.125$ &
$-0.989\pm 0.573$ & $-0.774\pm 0.204$ \\
  \\ \hline

\end{tabular}
\end{table*}

\end{document}